# ON THE QUANTITATIVE DETERMINATION OF HOLE-CONCENTRATION IN HIGH-TEMPERATURE CUPRATE SUPERCONDUCTORS

TATSUYA HONMA

*Department of Physics, Asahikawa Medical University,*
*Asahikawa, Hokkaido 078-8510, Japan*
*honma@asahikawa-med.ac.jp*

PEI-HERNG HOR

*Department of Physics and Texas Center for Superconductivity, University of Houston,*
*Houston, TX. 77204-5005, USA*
*phor@uh.edu*

We compared four hole-scales that have been used to determine the hole-concentration in high-temperature cuprate superconductors. We show that the hole-scale, $P_{pl}$-scale, based on the thermoelectric power [T. Honma *et al.*, Phys. Rev. **B70**, (2004) 214517.] is quantitatively consistent with spectroscopic probes for many different cuprate materials, while the other hole-scales, based on a well-known dome-shaped $T_c$-curve [M. R. Presland *et al.*, Physica **C176**, 95 (1991)], the *c*-axis lattice parameter [R. Liang *et al.*, Phys. Rev. **B73**, (2006) 180505(R).], and Hall coefficient [Y. Ando *et al.*, Phys. Rev. **B61**, (2000) 14956(R).], are not. We show that the quantitatively different hole-scales resulted in opposite conclusion of the same experimental observations. It can also lead to different interpretations of the electronic phase diagram when comparing different physical properties in different high-$T_c$ systems. We suggest that the $P_{pl}$-*scale* is the correct universal scale that works for all high-$T_c$ cuprates and it should be used for all quantitative doping dependence studies of cuprates.

## 1. Introduction

High-temperature cuprate superconductors (HTCS) have a very wide doping range that resulted in numerous doping dependent studies of various physical properties. While doping dependent studies of HTCS are becoming more and more quantitative through continuous improvements on experimental conditions, such as sample quality, experimental resolution and techniques, in the past twenty-eight years there has been, unfortunately, a misinterpretation of the corresponding data due to the use of a, although popular, quantitatively incorrect scale for determining the hole concentration. In this report, we address two issues of the doped-hole inhomogeneity and competition between magnetic order and spin density wave to demonstrate the fundamental importance of using the quantitatively correct scale to arrive at the physically correct

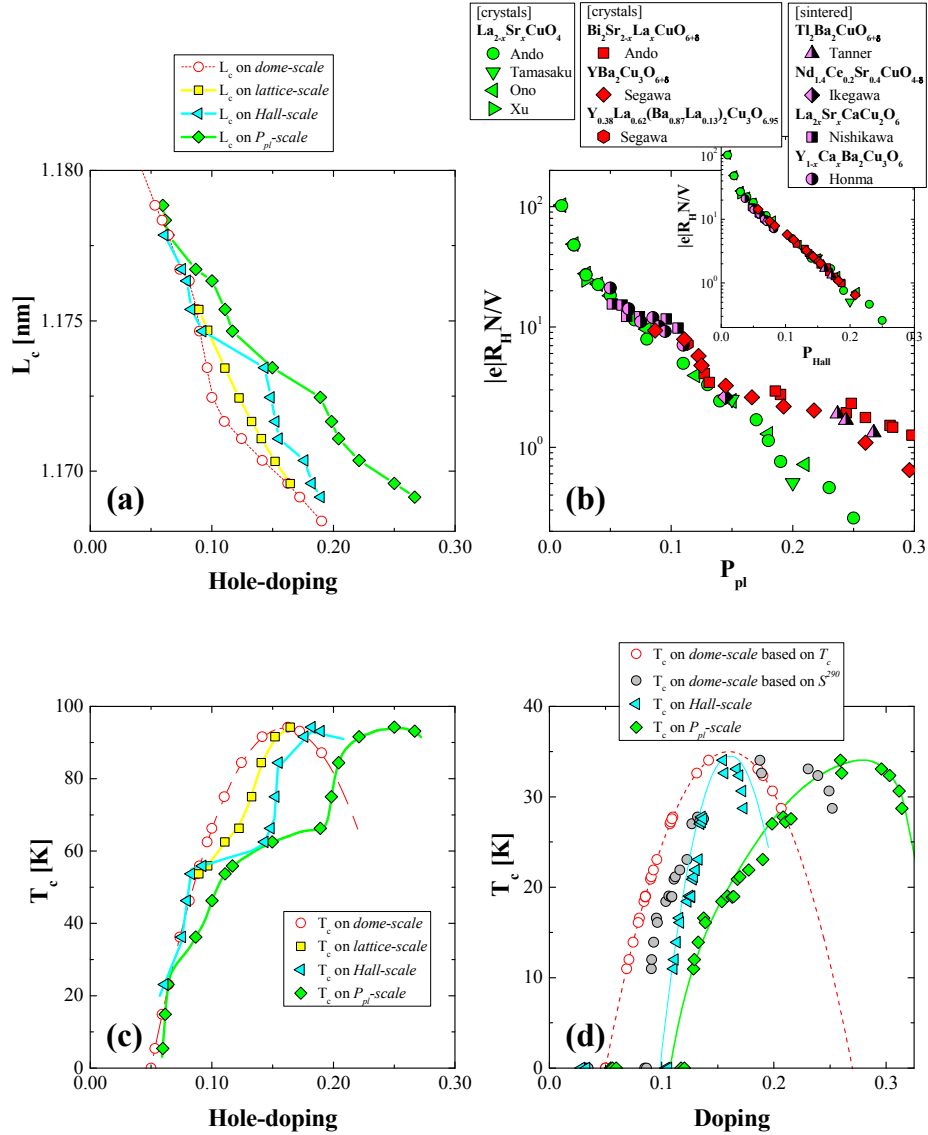

Fig. 1 (a) Doping dependence of $L_c$ as a function of different hole-scales. The plotted data were extracted from Ref. 7. (b) Doping dependence of $eR_HN/V$ at room temperature as a function of $P_{pl}$. Inset: $eR_HN/V$ at room temperature as a function of $P_{Hall}$. The plotted data were extracted from Refs. 11-22. All plotted data have both $R_H$ and $S^{290}$. (c) $T_c$ of $YBa_2Cu_3O_{6+\delta}$ versus different hole-scales. (d) $T_c$ of $Bi_2Sr_{2-z}La_zCuO_{6+\delta}$ versus different scales. The colored line is a guide to the eyes. The plotted data were extracted from Ref. 8.

conclusions and to address the subtle issues when comparing the doping dependence among different HTCS materials.

The most popular hole-scale is a measure of the hole concentration using the doping dependence of a superconducting transition temperature, $T_c$, that follows a dome-shaped $T_c$-curve with an empirical formula of $T_c = T_c^{max}\{1-82.6(P_{dome}-0.16)^2\}$, where $P_{dome}$ is the hole concentration, the value of 0.16 is the universal optimal hole concentration and $T_c^{max}$ is a maximum in $T_c$.[1] As seen in Fig. 1(c) where the $T_c$-curve is plotted as red circles connected by the dotted line. Instead of using $T_c$, the thermoelectric power at 290 K, $S^{290}$, was used to determine $P_{dome}$ by scaling the $S^{290}$ to the hole concentration based on the dome-shaped $T_c$-curve.[2,3] We will call the hole-scale based on the dome-shaped $T_c$-curve the "*dome-scale*" or $P_{dome}$-scale. The *dome-scale* is very convenient scale for estimating the hole concentration since $T_c$ is always available and it is still very popular now. However, the *dome-scale* has some self-inconsistency issues. For instance, in $La_{2-x}Sr_xCuO_4$, the $P_{dome}$ determined from $S^{290}$ is not corresponding to that from $T_c$.[4,5] Furthermore, in $YBa_2Cu_3O_{6+\delta}$, there is a well-known plateau around $T_c$ = 60 K and 90 K clearly observed in the oxygen-content dependence[6] which is completely washed out in the *dome-scale*. To address the missing plateaus a scale for $YBa_2Cu_3O_{6+\delta}$ based on the lattice parameter along *c*-axis, $L_c$, was proposed by Liang *et al*.[7] We will refer the hole concentration determined by the Liang's approach as $P_{lattice}$ and call it "*lattice-scale*". In Fig. 1(a), we plot the same $L_c$ data[7] of $YBa_2Cu_3O_{6+\delta}$ as a function of *doping* of the different scales. In the *dome-scale*, $L_c$ rapidly decreases at around 1/8 with doping. In the *lattice-scale* the rapid decreases become smoother. Ando *et al*. pointed out also in $Bi_2Sr_{2-x}La_xCuO_{6+\delta}$ that the $P_{dome}$ determined from $S^{290}$ is not corresponding to that from $T_c$.[8] They also proposed another scale based on the Hall coefficient, $R_H$.[8] The scale is based on the normalized $R_H$, $eR_HN/V$, where $V$ is a unit cell volume and $N$ is the $CuO_2$ layer number per the unit cell.[8] They tried to determine the hole concentration by scaling the temperature dependence of $eR_HN/V$ of many cuprates to that of $La_{2-x}Sr_xCuO_4$. Unfortunately, the scaling actually works only in the temperature range from 200 K to 300 K. Since in this temperature range, the temperature dependence of $R_H$ becomes very weaker,[4] the $eR_HN/V$ at room temperature can be representative of their scale. We will refer the hole concentration determined by the Ando's approach as $P_{Hall}$ and call it "*Hall-scale*". In the inset of Fig. 1(b), we plot the $eR_HN/V$ at RT as a function of $P_{Hall}$, which is equal to Sr-content in $La_{2-x}Sr_xCuO_4$. The $P_{Hall}$ is determined by scaling the value of $eR_HN/V$ to the doping concentration of $La_{2-x}Sr_xCuO_4$ as shown in the inset of Fig. 1(b). In Fig. 1(c), we plot the $T_c$ data[7] of $YBa_2Cu_3O_{6+\delta}$ as a function of *doping* of the different scales. According to the *lattice-scale*, as shown in Fig. 1(b), the dome-shaped $T_c$-curve has plateau at around 0.12. Essentially $T_c(P_{lattice})$-curve follows the dome-shaped $T_c$-curve but it is more refined such that an indication of the 60 K plateau is observable at ~0.12. Although $T_c(P_{Hall})$-curve tends to follow the $T_c(P_{dome})$-curve below 0.09, the $T_c(P_{Hall})$-curve shows a clear 60 K plateau and the optimal concentration appears at $P_{Hall}$ ~ 0.18.

We proposed a hole-scale which is also based on $S^{290}$ but quantitatively different from *dome-scale*.[5] Our proposed scale follows the experimental result that the hole concentration, $P_{pl}$, versus $S^{290}$ in $Y_{1-x}Ca_xBa_2Cu_3O_6$, where the $P_{pl}$ is equal to a half of the Ca-content, is corresponding to that in $La_{2-x}Sr_xCuO_4$, where the $P_{pl}$ is equal to the Sr-content.[5] It was also confirmed that the relation of $S^{290}$ versus $P_{pl}$ in the cation-doped HTCS is consistent with the hole concentration of the oxygen-doped HTCS, such as $HgBa_2CaCu_2O_{6+\delta}$ and $HgBa_2Ca_2Cu_3O_{8+\delta}$, and cation/oxygen co-doped HTCS, such as $Bi_2Sr_{2-x}La_xCuO_{6+\delta}$ and

$(Hg_{0.5}Fe_{0.5})Sr_2(Y_{1-x}Ca_x)Cu_2O_{7-\delta}$.[9] We will call our proposed hole-scale as the "$P_{pl}$-scale". One of the distinct features of $P_{pl}$-scale is that the optimal hole concentration, $P_{pl}^{opt}$, depends on the HTCS materials, it distributes in the doping range from 0.16 to 0.28 and centered around $0.24 \pm 0.01$.[9] This is quite distinct from the universal optimal doping concentration in the $P_{dome}$-scale. Further, for almost all HTCS, except $La_{2-x}Sr_xCuO_4$, the doping dependence of $T_c$ does not follow a dome-shaped $T_c$-curve. Almost all HTCS actually follows a half-dome-shaped $T_c$-curve.[9] In $YBa_2Cu_3O_{6+\delta}$, the doping dependence of $T_c$ follows a two-plateau $T_c$-curve as shown in Fig. 1(c).[10] The well-known 60 K plateau was clearly observed in $P_{pl}$-scale but absent in *dome-scale*.[10] It is clear that, from Fig. 1(a), *lattice-scale* is quantitatively different from $P_{pl}$-scale. From Fig. 1(b), *Hall-scale* is also quantitatively different from $P_{pl}$-scale. In Fig. 1(b), we plot the $eR_HN/V$ of $La_{2-x}Sr_xCuO_4$,[11-22] and the other materials as a function of $P_{pl}$. Although $eR_HN/V$ of the other materials roughly follows that of $La_{2-x}Sr_xCuO_4$ below $P_{pl} \sim 0.15$, $eR_HN/V$ of $YBa_2Cu_3O_{6+\delta}$, $Bi_2Sr_{2-x}La_xCuO_{6+\delta}$ and $Tl_2Ba_2CuO_{6+\delta}$ over $P_{pl} \sim 0.15$ deviates upward from that of $La_{2-x}Sr_xCuO_4$. In Fig. 1(d), we plot the $T_c$ of $Bi_2Sr_{2-z}La_zCuO_{6+\delta}$ against $P_{pl}$, $P_{dome}$ and $P_{Hall}$. The optimal doping and doping dependence in the $P_{pl}$-scale are quite different from those in *dome-scale* and *Hall-scale*.

In this report, we analyzed the doping dependent data of HTCS by using of $P_{pl}$-scale, and compared our conclusions with that derived from other hole-scales. $P_{pl}$ is determined by comparing the value of $S^{290}$ data with $P_{pl}$-scale, which is more reliable.[5,9] As the second method, for $YBa_2Cu_3O_{6+\delta}$, $P_{pl}$ is determined by comparing the value of $T_c$ with the two-plateau $T_c(P_{pl})$-curve as shown in Fig. 1(c).[10] For $HgBa_2CuO_{4+\delta}$, the value of $P_{pl}$ is determined by comparing the corresponding value of $T_c$ in the paper with $T_c$-curve determined from $S^{290}$ data[23] by using $P_{pl}$-scale. As the third method, we can determine $P_{pl}$ from either the in-plane conductivity, $\sigma_{ab}$, or out-of-plane conductivity, $\sigma_c$, since there is a universal doping dependence of $\sigma_{ab}$[24] and $\sigma_c$[25] on $P_{pl}$. We always selected the paper that reports the value of $S^{290}$ first and used the data with the value of $T_c$ second. We also report $R_H$ of $Y_{1-x}Ca_xBa_2Cu_3O_6$ with no chain. The details of sample preparation were reported in Refs. 5,9,26. The Hall effect was measured by dc method under a magnetic field up to 7 T[26] or by physical property measurement system (PPMS, Quantum Design). Both results were consistent. In analyzing the reported data by $P_{pl}$-scales, we selected the data with either $S^{290}$ or $T_c$ reported in the literature.

## 2. Quantitative comparison of $P_{pl}$-scale with the other working hole-scales

In Fig. 2, we plot the hole concentration determined by many spectroscopic probes as a function of $P_{pl}$.[27-29] For comparison we also plot the corresponding hole concentration of typical HTCS determined by the *dome-scale* and the *lattice-scale* as color-coded lines and dotted line, respectively. The hole concentration measured by near edge x-ray absorption fine structure, NEXAFS,[27] and nuclear quadruple resonance, NQR,[29] are consistent with $P_{pl}$ within a error band of $\pm 0.01$, consistent with the uncertainty of the $P_{pl}$-scale.[5,9] Further, we also plot $P_{Hall}$ of various HTCS materials.[11-22] While $P_{pl}$ corresponds well to the hole concentration determined by NEXAFS and NQR. In contrast, the $P_{dome}$, $P_{lattice}$ and $P_{Hall}$ scales are clearly deviate substantially from $P_{pl}$.

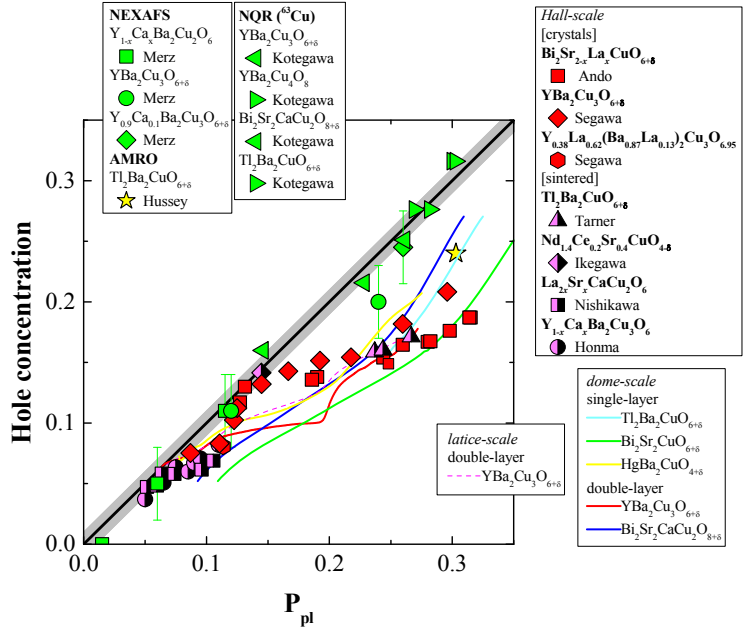

Fig. 2 $P_{pl}$ versus the hole concentration measured by various experimental probes. The plotted data were extracted from Refs. 27-29. For Hall-scale, $P_{pl}$, except of $Tl_2Ba_2CuO_{6+\delta}$, was estimated from $S^{290}$ data. The other $P_{pl}$ were determined from $T_c$ data.

Here, we demonstrate the first example that completely opposite conclusions resulted when using the $P_{pl}$-*scale* and $P_{dome}$-scale to analyze data. While many HTCS are shown to be having an inhomogeneous doped-hole concentration[44,45] the absence of doped-hole inhomogeneity in $YBa_2Cu_3O_{6+\delta}$ seems to be the only exception. Bobroff *et al.*[34] reported that the NQR spectrum of $YBa_2Cu_3O_{6+\delta}$ showed that the inhomogeneity in the doping distribution is quite small. However we need to point out that all the conclusions in Ref. 34 were based on the $P_{dome}$-*scale*.

We first compare the doped-hole concentration determined by nuclear magnetic resonance, NMR, using $P_{pl}$-*scale* and $P_{dome}$-scale. In Fig. 3, we plot $^{89}Y$ Knight shift, $^{89}K_s$, at room temperature, RT, of Y-based HTCS, as a function of $P_{dome}$ in the top panel and same $^{89}K_s$ as a function of $P_{pl}$ in the lower panel. In Fig. 3(a), we also draw the two lines with the linear slopes of 580 ppm/hole and 823 ppm/hole reported in Refs. 30,31. While each linear slope may represent the limited data selected in the corresponding paper, it clearly showed that the doping dependence of $^{89}K_s$ cannot be represented by a straight line using $P_{dome}$-scale when we use all available $^{89}K_s$ at RT in Fig. 3(a). This suggests a large ambiguity of using $^{89}K_s$ data to determine hole concentration based on the $P_{dome}$-scale. However, in the $P_{pl}$-scale as shown in Fig. 3(b), all $^{89}K_s$ at RT lie on one straight line represented by $555P_{pl} - 68$. The deviation at $P_{pl} < 0.07$ may come from the AF phase.[32,33] The $^{89}K_s$ increases linearly with $P_{pl}$ within the error bar of the $P_{pl}$-scale. However $^{89}K_s$ is barely linearly proportional to $P_{dome}$ with a large scattering.

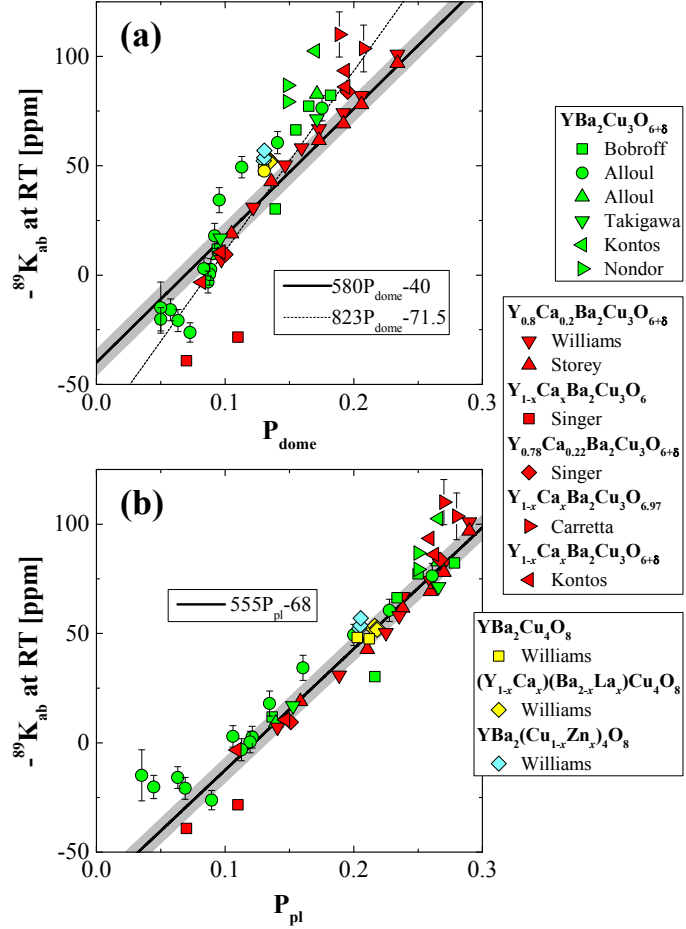

Fig. 3 The doping dependence of the $^{89}$Y Knight shift ($^{89}K_s$) at room temperature (RT) for Y-based HTCS. (a) $^{89}K_s$ at RT versus $P_{dome}$. (b) $^{89}K_s$ at RT versus $P_{pl}$. Both plotted data are from the same data set. The only difference is the hole-scales used to analyze the data set. The plotted data are from Refs. 30-43.

In Ref. 34, from $^{89}$Y NMR Fourier transform spectra at 300 K, the doping distribution, $\Delta P_{dome}$, of the slightly overdoped YBa$_2$Cu$_3$O$_7$ and the underdoped YBa$_2$Cu$_3$O$_{6.6}$ were deduced by using the relation of the slope of 580 ppm/hole based on the $P_{dome}$-scale.[30] The full width at half maximum (FWHM) of $^{89}Y$ NMR spectra for $YBa_2Cu_3O_{6+\delta}$ led to a conclusion that the doping distribution $\Delta P_{dome}$ < 0.025 for $YBa_2Cu_3O_7$ and $\Delta P_{dome}$ < 0.01 for $YBa_2Cu_3O_{6.6}$ as shown in Fig. 4(a). These values are much smaller than $\Delta P_{dome}$ = 0.1 in $La_{2-x}Sr_xCuO_4$[44] and $Bi_2Sr_2CaCu_2O_{8+\delta}$.[45] However if we calculate the doping distribution using $P_{pl}$-scale, $\Delta P_{pl}$, from their FWHM and the relation of $555P_{pl}$ - 68, the $\Delta P_{pl}$, as seen Fig. 4(b), of $YBa_2Cu_3O_{6.6}$ is 0.07 and that of $YBa_2Cu_3O_7$ is ~0.03. We reproduce the doping distribution and $T_c$ of Fig. 2 of Bobroff *et al*.[34] in Fig. 4(a) and re-plot the same data set by directly converting their hole concentration to $P_{pl}$ in Fig. 4(b). Therefore, as seen in Figs. 4(a) and 4(b), using the **SAME** NMR data, the electronic spread determined by $P_{pl}$-scale is seven times and one and a half as large as those determined by the $P_{dome}$-scale for underdoped and slightly overdoped $YBa_2Cu_3O_{6+\delta}$, respectively. The doping distribution estimated by the $P_{pl}$-*scale* is comparable to other cuprates. Therefore not only our conclusion is opposite to that when using $P_{dome}$-scale but also more importantly, based on all currently available data, implies that doped-hole inhomogeneity is a generic property of all HTCS.

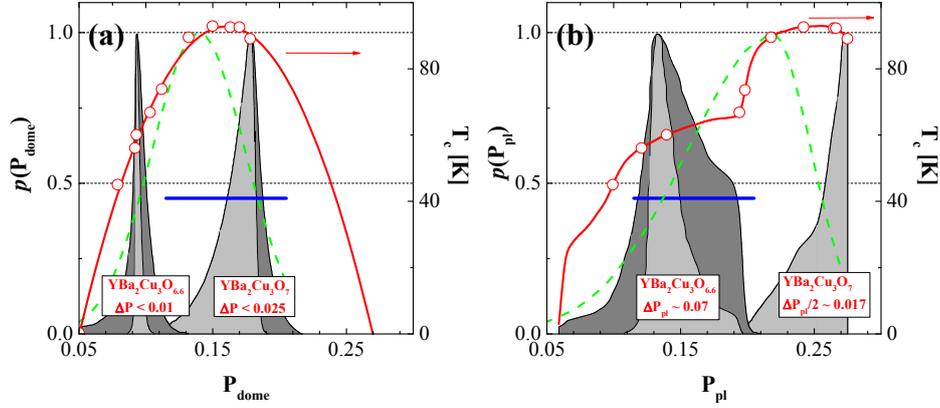

Fig. 4 Doping dependence of $T_c$ and distribution of hole concentration for $YBa_2Cu_3O_{6+\delta}$. (a) The reproduction of the lower panel of Fig. 1 in Bobroff's paper.[34] The electronic inhomogeneity was calculated by using the $P_{dome}$-scale for the $YBa_2Cu_3O_{6.6}$ ($T_c$ = 60 K phase) and $YBa_2Cu_3O_7$ ($T_c$ = 90 K phase). The red line shows the well-known superconducting dome, with $T_c^{max}$ = 93 K, reported in Ref. 9. The dark and light gray are corresponding to those in Fig. 1 in Ref. 34. The dashed distribution and the line represent the hole distributions of $La_{2-x}Sr_xCuO_4$[44] and $Bi_2Sr_2CaCu_2O_{8+\delta}$.[45] (b) We re-plot Fig. 1(a) by converting $P_{dome}$ into $P_{pl}$. The red line shows the double plateau $T_c$-curve, with $T_c^{max}$ = 93 K, reported in Ref. 9.

As a second example, we demonstrate the universal nature of our quantitative scale by focusing on the translational symmetry preserving magnetic order ($T_{mag}$) recently observed in $YBa_2Cu_3O_{6+\delta}$ by polarized neutron in the pseudogap state.[46] In the heavily underdoped regime where incommensurate spin-density-wave, SDW, order ($T_{SDW}$) exists in the $YBa_2Cu_3O_{6+\delta}$, the $T_{mag}$ seems to be much reduced, suggesting that they are competing with each other.[47,48] Same magnetic order is also observed in the $HgBa_2CuO_{4+\delta}$,[49] although the SDW has not been reported in the same system yet. Depending on how the hole concentration is estimated, the competition between magnetic order and SDW cannot be pinned down in the $HgBa_2CuO_{4+\delta}$.[50] In Fig. 5(a), we reproduce the magnetic phase diagram of Fig. 6 in Ref. 50, $YBa_2Cu_3O_{6+\delta}$,[46-48,51] $Y_{0.85}Ca_{0.15}Ba_2Cu_3O_{6+\delta}$,[46] and $HgBa_2CuO_{4+\delta}$.[49,50] In Ref. 46, the data of $YBa_2Cu_3O_{6+\delta}$ and $Y_{0.85}Ca_{0.15}Ba_2Cu_3O_{6+\delta}$ are plotted as a function of $P_{lattice}$ and that of $HgBa_2CuO_{4+\delta}$ is plotted as a function of $P_{dome}$, determined from the $S^{290}$ data[9] according *dome-scale*. The different scales used by different groups present difficulties to compare experimental observations from different group and, furthermore, add substantial confusions in case a conclusion was drawn without properly address the issue of using a quantitatively consistent hole-scale. To evaluate the relation between $T_{mag}$ and $T_{SDW}$, we also plot $T_{SDW}$ reported in Ref. 48 into Fig. 5(a). According to Ref. 50 if the Fig. 5(a) is correct, it would necessarily imply that a competition between the SDW and the $q = 0$ magnetic order is either absent in $HgBa_2CuO_{4+\delta}$ or will only commence at much lower doping than that in $YBa_2Cu_3O_{6+\delta}$. On the other hand, if the inset of Fig. 5(a) is correct, the $q = 0$ order in $HgBa_2CuO_{4+\delta}$ has not yet been investigated to a hole concentration as low as $P_{lattice} \sim 0.073$ of $YBa_2Cu_3O_{6+\delta}$, and no conclusion can be drawn at this point regarding its competition with the SDW order. In Fig. 5(b), we plot the same data set as a function of the reduced doped-hole concentration $P_{pl}/P_{pl}^{opt}$, and compare them with universal electronic phase diagram (UEPD) reported in Ref. 9 based on the $P_{pl}$-scale. According to the $P_{pl}$-scale, while $T_{mag}$ of both $YBa_2Cu_3O_{6+\delta}$ and $HgBa_2CuO_{4+\delta}$ occurs at the upper pseudogap temperature for $P_{pl}/P_{pl}^{opt} > 0.4$. But, the $T_{mag}$ of $YBa_2Cu_3O_{6+\delta}$ rapidly decreases with undoping over $P_{pl}/P_{pl}^{opt} \sim 0.4$ where SDW appears, although there is no data of $HgBa_2CuO_{4+\delta}$ with $P_{pl}/P_{pl}^{opt} < 0.4$. Regarding the question in Ref. 50, our answer is that $P_{pl}$ of $HgBa_2CuO_{4+\delta}$ investigated is still too high to be compare with $YBa_2Cu_3O_{6+\delta}$ for addressing the competition between $T_{mag}$ and $T_{SDW}$. More importantly, it is clear that the magnetic phase diagram of $HgBa_2CuO_{4+\delta}$ is essentially the same as that of $YBa_2Cu_3O_{6+\delta}$ for $P_{pl}/P_{pl}^{opt} > 0.4$ and translational symmetry preserving magnetic order, $T_{mag}$, is intimately related to the upper pseudogap reported in our UEPD.

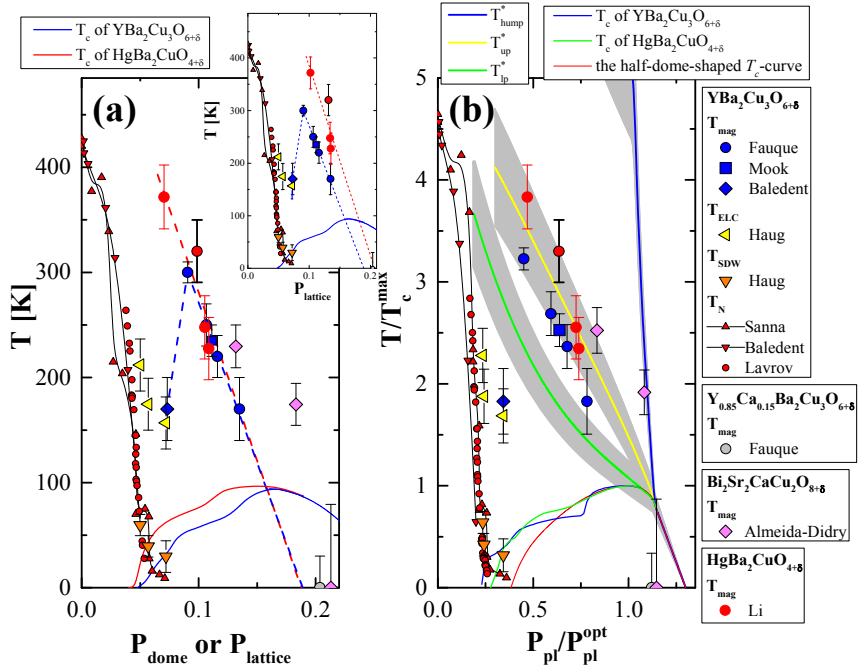

Fig. 5 Magnetic phase diagram of $YBa_2Cu_3O_{6+\delta}$, $Y_{1-x}Ca_xBa_2Cu_3O_{6+\delta}$, $Bi_2Sr_2CaCu_2O_{8+\delta}$ and $HgBa_2CuO_{4+\delta}$. The plotted data were extracted from Refs. 46-51. (a) Main figure and the inset are corresponding to those of Fig. 6 in Ref. 47, respectively. The carrier concentration for $YBa_2Cu_3O_{6+\delta}$ is based on *lattice-scale*.[7] The carrier concentration for $HgBa_2CuO_{4+\delta}$ is determined from $S^{290}$ data[23] according to *dome-scale* in the main figure and from the $T_c(P_{lattice})$-curve of $YBa_2Cu_3O_{6+\delta}$ in the inset. (b) All carrier concentration is based on the $P_{pl}$-scale. $P_{pl}^{opt}$ is 0.25 for $YBa_2Cu_3O_{6+\delta}$ and 0.235 for $HgBa_2CuO_{4+\delta}$.[9,10] $T_N$ of $YBa_2Cu_3O_{6+\delta}$ are coming from Refs. 47,53,54.

Finally, we want to point out some recent theoretical approaches to the *hole-scale*.[55-57] The $P_{pl}$-scale is based on a universal doping dependence of $S^{290}$.[5,9] Although the *dome-scale* is based on the universal optimal hole-concentration of 0.16,[1] it demonstrates another universal doping dependence of $S^{290}$, except of $La_{2-x}Sr_xCuO_4$.[2] In order to compare the theoretical works with the *hole-scales*, it is convenient to use a characteristic doped-hole concentration, $P_s$, where a sign change of $S^{290}$ occurred. Experimentally, the value of $P_s$ is universally ~0.25 in the $P_{pl}$-scale,[5,9] but ~0.18 in the *dome-scale*.[2,4] Theoretically, the doping dependence of $S^{290}$ is reproduced within the Hubbard model with the moderate on-site repulsion[55] and using a simplified Hubbard model of a bad metal.[56] In both cases $P_s \sim$ 0.2. Furthermore, the doping dependence of $S^{290}$ at 0 K, including the sign change, is attributed to a possible underlying quantum critical point.[57] Unfortunately neither could distinguish the *dome-scale* from $P_{pl}$-scale.

## 3. Conclusions

In summary, we have demonstrated the quantitative consistency and the use of the universal $P_{pl}$-scale. Independent of any theoretical considerations we shall emphasize that $P_{pl}$-scale is consistent with the carrier concentration estimated by many different experimental probes in many cuprates. We suggest that the $P_{pl}$-scale should be used for all quantitative doping dependent studies of HTCS.